\documentstyle[12pt,a4]{article}
\begin{document}
\def\doublespaced{\baselineskip=\normalbaselineskip\multiply\baselineskip
  by 150\divide\baselineskip by 100}
\doublespaced
\def\lsim{~{\rlap{\lower 3.5pt\hbox{$\mathchar\sim$}}\raise 1pt\hbox{$<$}}\,}
\def\gsim{~{\rlap{\lower 3.5pt\hbox{$\mathchar\sim$}}\raise 1pt\hbox{$>$}}\,}
\def\thisday{~\today ~and~ hep-ph/0207137}


\begin{titlepage}
\vspace{0.5cm}
\begin{flushright}
hep-ph/0207137    \\
July 2002
\end{flushright}
\vspace{0.5cm}
\begin{center}
\large
{Neutrino Oscillations Due to Nonuniversal Gauge Interactions in 
the Weak Sector}
\end{center}
\begin{center}
{\bf Ehab Malkawi\footnote{
e-mail:malkawie@just.edu.jo}}\\
{Department of Physics,
Jordan University of Science \& Technology\\
 Irbid 22110, Jordan}\\
\end{center}
\vspace{0.4cm}
\raggedbottom
\relax
\begin{abstract}
\noindent
We discuss the structure of the neutrino 
mass matrix as derived from a nonuniversal
electroweak gauge interaction model. We discuss two interesting 
patterns of neutrino masses.
The first pattern is hierarchal which fits the LMA or LOW solutions of 
solar neutrino data. The second pattern gives rise to inverted mass 
spectrum which fits the currently not preferred SMA solution. The mechanism for 
generating a light sterile neutrino mass 
is interesting and is briefly discussed.
\end{abstract}
\vspace{0.5cm}
\vspace{1.0cm}
\end{titlepage}

\section*{I Introduction}
 
New results from several experiments on neutrino physics 
\cite{solar,kam,chooz,lsnd} 
have a convincing evidence of neutrino masses and mixings.
A large body of information on the neutrino mass matrix structure has been 
established from solar \cite{solar}, atmospheric \cite{kam}, 
reactor \cite{chooz}, and accelerator \cite{lsnd} experiments. 
Still, the complete picture of the mass matrix, as derived 
from neutrino data, is far from complete. 
Further data are needed to constrain the large number of parameters 
in the mass matrix. Results on neutrinoless double-beta decay \cite{double},
if confirmed,  
provide further information that cannot be provided by oscillation 
experiments.  

Extensions of the Standard Model (SM) and/or new symmetries, e.g., horizontal
symmetries, have been discussed thoroughly in connection with 
neutrino oscillation data \cite{alter,bar1,cpt,4nu,nuexd,stab}.
A possibility to explain neutrino data through 
non-universal gauge interactions in the electroweak sector was discussed 
in Ref.~\cite{ehab2}, where the third 
generation of fermions are assumed to transform under a weak $SU(2)_h$ 
symmetry, while the first and second generations transform 
under a different $SU(2)_l$. 
The phenomenology of the scenario and similar versions were extensively
 discussed in Refs.~\cite{ehab2,ehab1} and earlier in Ref.~\cite{lima}. 
The basic idea is the assumption of a nonuniversal electroweak gauge  
symmetry $SU(2)_l\times SU(2)_h\times U(1)_Y$, where 
different generations of fermions transform under different $SU(2)$ 
symmetries.
The scenario is phenomenologically well motivated and consistent with 
all low energy data as long as the heavy gauge boson masses are at least of the
order of 1 TeV \cite{ehab1}. 
However, as 
pointed out in Ref.~\cite{ehab2}, 
the scenario fails to give rise to a maximal mixing 
in order to account for the atmospheric data. This is due to 
the fact that the second and third families
can only mix weakly through the ratio of the two vacuum expectation values
of the two Higgs doublets.
In this work, we modify the previous scenario by assuming that
 the second and third generations 
of fermions transform under $SU(2)_h$, while the first generation
transform under $SU(2)_l$. In this case we can produce maximal
mixing in the atmospheric sector and be consistent with other neutrino data.  

It could be argued that the phenomenology of the new scenario is less 
interesting. This could be due to the precise low energy 
data on universality of the gauge interactions, e.g., regarding the
electron and muon interactions \cite{data}. This will simply have 
the effect of
pushing the mass of the extra heavy gauge bosons further up. 
Still, a detailed study of the
phenomenology of the model is highly needed to constrain the heavy gauge
boson masses and couplings. Nevertheless,  
the scenario can survive all low energy data
constraints provided the extra gauge bosons are heavy enough. 
On the other hand, the Yukawa 
sector is not constrained by low energy data and this is the motive 
for considering such a scenario. A study of 
the neutrino mass matrix is performed in this work independent of the
gauge sector and low energy data.

 The rest of the paper is organized as follows: 
 In Sec. II, we briefly review the model and then extract the general form
the mass matrix. In Sec. III, we discuss the structure of the mass matrix
and discuss two possible patterns of neutrino masses. We also 
discuss briefly the possibility of including a light sterile neutrino.

\section*{II Structure of the model}

The scenario we discuss is   
based on the gauge symmetry G=~$SU(3)_c\times SU(2)_l\times
SU(2)_h\times U(1)_Y$. Where, the left-handed 
second and third fermion generations are subjected to a weak  
gauge interaction described by $SU(2)_h$. On the contrary, the first
generation is subjected to another $SU(2)_l$ gauge interaction.
The $U(1)_Y$ group is the SM
hypercharge group. The right-handed fermions only transform under the 
$U(1)_Y $ group as assigned by the SM. Finally, the QCD interactions and the
color symmetry $SU(3)_c$ are the same as that in the SM.

The spontaneous symmetry-breaking of the group G
is accomplished by introducing the
complex scalar fields $\Sigma$, $\Phi_1$, and $\Phi_2$, where 
$\Sigma\sim (1,2,2,0) $, $ \Phi_1 \sim (1,2,1,1)$, and $\Phi_2\sim (1,1,2,1)$.
The group G is then broken at three different stages. 
The first stage of symmetry breaking is accomplished once the $\Sigma $
field acquires a vacuum expectation value (vev) $u$, i.e.,  
$
\left\langle \Sigma \right\rangle =\pmatrix{u & 0 \cr 0 & u \cr}\, ,
$
where $u$ is expected to be of the order of 1 TeV.
The form of $\left\langle \Sigma \right\rangle$  
guarantees the breakdown of $SU(2)_l \times SU(2)_h\rightarrow SU(2)$.
Therefore, the unbroken symmetry  
is essentially the SM gauge symmetry $SU(3)_c\times SU(2)\times U(1)_Y$, 
where $SU(2)$ is the usual SM weak group. 
At this stage,
three of the gauge bosons acquire a mass of order $u$, while the other
gauge bosons remain massless. The second and third stage of 
symmetry breaking (the electroweak symmetry-breaking) is  accomplished
through the scalar fields $\Phi_1$ and $\Phi_2$ by
acquiring their vacuum expectation values  
$
\left\langle \Phi_1 \right\rangle=\pmatrix{0 \cr v_1\cr}\,
$,
and $
\left\langle \Phi_2 \right\rangle=\pmatrix{0 \cr v_2\cr}\,
$, respectively. The electroweak symmetry-breaking scale $v$ is defined 
 as $v\equiv\sqrt{v_1^2+v_2^2}=246$ GeV.
Since the third fermion generation  
is heavier than the first generation, it
is suggestive to conclude that $v_2\gg v_1$.

Fermions
acquire their masses through Yukawa interactions via the 
$\Phi_1 $ and $\Phi_2$ scalar fields. 
The full neutrino Yukawa interaction terms are given by the Lagrangian 
\begin{eqnarray}
 {\cal L}_{\mbox{\rm{Yukawa}}} &=&
       \overline{\Psi _L}^1{\widetilde{\Phi}}_1 \left[
       g_{11}^\nu \nu_{eR} +g_{12}^\nu \nu_{\mu R} + 
       g_{13}^\nu \nu_{\tau R}\right]+\nonumber \\
 &&\ \overline{\Psi _L}^2 {\widetilde{\Phi}}_2 \left[ g_{21}^\nu \nu_{eR}+g_{22}^\nu \nu_{\mu R} +
    g_{23}^\nu\nu_{\tau R} \right] +\nonumber \\
&&\ \overline{\Psi _L}^3 {\widetilde{\Phi}}_2 \left[
 g_{31}^\nu \nu_{eR}+g_{32}^\nu \nu_{\mu R}+g_{33}^\nu\nu_{\tau R} 
\right] +h.c., 
\label{dirac}
 \end{eqnarray}
where ${\widetilde{\Phi}}_{1,2}\equiv i\tau_2 \Phi_{1,2}^*$ and 
where 
\begin{equation}
\Psi_L^1=\pmatrix{\nu_{eL}\cr e_L}\, ,\, 
\Psi_L^2=\pmatrix{\nu_{\mu L}\cr \mu_L}\, , \, \rm{and}\,\, 
\Psi_L^3=\pmatrix{\nu_{\tau L}\cr \tau_L}\, .
\end{equation}

The Dirac mass matrix derived from Eq.~(\ref{dirac}) is written as
\begin{equation}
M_D=\pmatrix{g_{11}^\nu v_1 & g_{12}^\nu v_1 & g_{13}^\nu v_1 \cr 
             g_{21}^\nu v_2 & g_{22}^\nu v_2 & g_{23}^\nu v_2 \cr 
             g_{31}^\nu v_2 & g_{32}^\nu v_2 & g_{33}^\nu v_2 \cr} \, .
\label{full}
\end{equation}  
The right-handed neutrino Majorana mass matrix $M_R$ is assumed to have a common 
mass scale of the order of the GUT scale, $M_X\sim 10^{15}$ GeV.
Therefore, the full neutrino mass matrix forms a $8\times 8$ matrix which can be written as
\begin{equation}
M_\nu=\pmatrix{0& M_D\cr M_D^T & M_R\cr}\, .
\end{equation}
By invoking the seesaw mechanism the left-handed neutrino Majorana mass matrix is then given 
as
\begin{equation}
M_L= M_D M_R^{-1} M_D^T\, .
\label{ml}
\end{equation} 
Due to the seesaw mechanism all elements of $M_L$ are highly suppressed by the GUT scale $M_X$
of the right-handed Majorana mass matrix $M_R$.  
In this work we assume the charged lepton mass matrix is diagonal. 
In the next section we give 
further discussion of the derived neutrino mass matrix.  

\section* {III Neutrino masses and mixings} 

The most general form of the Majorana  mass matrix as given in Eq.~({\ref{ml}})
can be written as
\begin{equation}
M_L=m
\pmatrix{    g_{11} \epsilon^2 & g_{12} \epsilon & g_{13} \epsilon \cr 
             g_{12} \epsilon & g_{22} & g_{23}  \cr 
             g_{13} \epsilon & g_{23} & 1 \cr} \, .
\label{full2}
\end{equation} 
where, $\epsilon\equiv {v_1}/{v_2}$ and all other parameters 
inside the mass matrix are assumed of order 1. 
It is highly desirable to investigate what symmetries, discrete or continuous, 
could lead to a pattern that is consistent with the neutrino data. 
However, it is our intention in this work
to discuss the mass matrix in a general way without regard to 
the underlying symmetry.
There are two special forms of the mass matrix that could be compatible 
with data. The first form gives rise to a degenerate pattern in the 
neutrino masses and can be written as
\begin{equation}
M_L=m\pmatrix{ \epsilon^2 & \epsilon &  \epsilon \cr 
              \epsilon & 1 & 1  \cr 
             \epsilon & 1 & 1 \cr} \, .
\label{full4}
\end{equation} 
where coefficients of order 1 multiplying the small parameter $\epsilon$ has
been omitted. The mass matrix form has been discussed in 
Refs.~\cite{alter,bar1}
in connection with other scenarios.
In this case one finds  
two light eigenstates approximately given by $\pm\, m\epsilon$ and a heavy
mass eigenstate approximately given by $2m$. Also one finds
$\tan\theta_{23}\approx 1$,  
$\tan\theta_{12} \approx 1-\epsilon$, $\Delta m^2_{\rm{atm}}\approx 4m^2$, 
$\Delta m^2_{\rm{sol}} \approx m^2 \epsilon^2$, 
and $\theta_{13}\approx \epsilon$. 
A large value of $\epsilon$ is needed to recover the LMA solution
\cite{smir} which raises doubts to the perturbation approach. While, the
result could fit the LOW solution. 
For example, taking $\epsilon=0.1$ and other parameters around 1, we
can easily recover the best fit of the LOW solution, while a  
 $3\sigma$ is needed for the LMA solution. 
On the other hand, 
$m_{ee}\approx \Delta m^2_{\rm{sol}}$ is beyond the
reach of the next generation of experiments.

Next we consider a second form of the neutrino mass matrix 
which gives rise to
what is called the inverted pattern, where we get 
two heavy 
eigenstates with masses of 
order $O(v_2^2/M_X)$ and a third light mass eigenstate of 
order $O(v_1^2/M_X)$. In order for the scenario to 
give the correct pattern of mass splittings, 
we impose the condition that the two heavy states are 
degenerate within a small gap to explain the solar anomaly. While their 
large splitting with the light state explains the atmospheric anomaly. 
Thus, we consider the special form
\begin{equation}
M_L=m
\pmatrix{     \epsilon^2 &  \epsilon &  \epsilon \cr 
              \epsilon & -1 & a  \cr
              \epsilon & a & 1 \cr} \, .
\label{full3}
\end{equation}
where the parameter $a$ is written explicitly in the mass matrix 
and we omitted the order 1 coefficients multiplying $\epsilon$. 
The mass eigenstates can be readily determined as
\begin{equation}
m_1 \approx m\, \epsilon^2 \, , 
\end{equation}
\begin{equation}
m_2 \approx  -m\sqrt{1+a^2}+ O(\epsilon^2) \, ,
\end{equation}
\begin{equation}
m_3\approx  m\sqrt{1+a^2}+O(\epsilon^2)\, .
\end{equation}
The atmospheric scale, $\Delta_{\rm{atm}}$, is then associated with the 
mass splitting $\Delta_{13}$ (or $\Delta_{\rm{12}}$), where 
\begin{equation}
\Delta_{\rm{atm}}= \Delta_{13}=\Delta_{12}\approx m^2 \left(1+a^2\right)
\end{equation}
While the solar mass scale, $\Delta_{\rm{sol}}$, is associated 
with the mass splitting
$\Delta_{23}$, where
\begin{equation}
\Delta_{\rm{sol}}=\Delta_{23}\approx m^2\, \epsilon^2\, .
\end{equation}
For $a\approx 1$ we find that
$\epsilon\approx 0.1$ is consistent with results on mass splittings.
For the mixing angles we find 
We find that 
\begin{equation}
\tan\theta_{23}\approx \frac{a}{1+\sqrt{1+a^2}}\, .
\end{equation}
Note that for large $a$
\begin{equation}
\lim_{a\rightarrow \infty} \tan\theta_{23} =1\, .
\end{equation}
In fact for $\sin^2 2\theta_{23} \geq 0.7$, as indicated
by data, we require $a\geq 1.5$. 
Hence, the scenario can account for the large mixing angle needed 
to explain the atmospheric data.
For the mixing angle $\theta_{13}$ we find 
$\theta_{13} \approx  \epsilon$ which is also consistent 
with the CHOOZ bound \cite{chooz}.
For $a= 1.5$ and $\epsilon\approx 0.1$, we find that 
$\tan\theta_{13} \approx 0.1$. 
Finally, for the mixing angle $\theta_{12}$ we find
$\theta_{12}\approx \epsilon$ 
with the limit 
\begin{equation}
\lim_{a\rightarrow \infty}\tan\theta_{12} = \frac{\epsilon}{\sqrt{2}a}\, .
\end{equation}
The result gives rise to the SMA solution
of the solar neutrino data. For $a=1.5$ and $\epsilon\approx 0.1$, 
we find that $\sin^2 2\theta_{12}\approx 2\times 10^{-3}$. Recent global
 analysis of
of all neutrino data disfavors the SMA solution \cite{smir}. However,
more data is still needed to confirm such a result. If the SMA solution
turns out to be excluded with high confidence 
then the inverted pattern as derived from this
 scenario would be discarded because it can only lead to the
SMA solution of the solar neutrino data.

The scenario offers an interesting mechanism for embedding 
an extra light sterile neutrino which has been discussed in detail in
Ref.~\cite{ehab2}. In this case, one introduces  
an exotic bi-doublet fermion under the weak symmetries  
$SU(2)_{l,h}$. Once the first symmetry breaking is invoked, 
the bi-doublet fermion is split
into two pieces; 
a singlet which is identified as the sterile neutrino,  
and an active triplet field which can acquire a heavy Dirac mass 
in order to decouple from the low-energy regime.
Once the scalar field $\Sigma$ acquires its vev $u\gsim$ 1 TeV, 
the Dirac masses of the triplet active fermions and the sterile neutrino 
are generated of the same order 1 TeV \cite{ehab2}.
The mass matrix $M_D$ in Eq.~(\ref{full}) is then enlarged to a 4x4 matrix
with an extra parameter $u\gg v_2\gg v_1$. 
By invoking the see-saw mechanism, 
the mass matrix $M_L$ gives rise to a light 
sterile neutrino with mass of the order ${u^2}/{M_X}$. 
If, for example, we choose $u\approx 1$ TeV and 
$M_X \approx  10^{15}$ GeV, the mass of the sterile 
neutrino is of order 1 eV
which is compatible with LSND data \cite{lsnd}. 
This case has been discussed extensively 
in Ref.~\cite{ehab2} which is similar to the case we have.
The masses of the active and sterile neutrinos are controlled 
by the vacuum expectation values $v_1$, $v_2$, and $u$. For example, 
the ratio of the sterile neutrino mass to the atmospheric neutrino mass
is of the order $u^2/v_2^2 \approx 16 $, for $v_2\approx 250$ GeV and 
$u\approx 1$ TeV. Thus, it is unlikely that the model  
can generate a heavy sterile neutrino mass of order 1-10 keV 
as an interesting dark matter candidate \cite{dark}. 

A final comment is that the flavor changing neutral currents (FCNC) 
are not expected to be significant in our scenario 
due to the tight  constraints driven 
by low energy data \cite{ehab2}.  



\begin{thebibliography}{99}

\bibitem{solar}
T.A. Kirsten et al. (GALLEX and GNO Collaborations),
Nucl. Phys. Proc. Suppl. {\bf 77}, 26 (1999);
V.N. Gavrin et al. (SAGE COLLABORATION), 
Part. Nucl. Lett. {\bf 108}, 18 (2001);
B.T. Cleveland et al. (Homestake Collaboration), 
Astrophys. J. {\bf 496}, 505 (1998);
Y. Fukuda et. al., Phys. Rev. Lett. {\bf 86}, 5656 (2001);
Q.A. Ahmad et al. (SNO Collaboration), Phys. Rev. Lett. {\bf 87}, 071301 (2001).

\bibitem{kam}
Y. Fukuda et al. (Super-Kamiokande Collaboration),
Phys. Rev. Lett. {\bf 85}, 3999 (2000);
G. Giacomelli et al. (Macro Collaboration), 
hep-ex/0110021.

\bibitem{chooz}
C. Bemporad et al. (CHOOZ collaboration), 
Nucl. Phys. Proc. Suppl. {\bf 77}, 159 (1999).

\bibitem{lsnd} G.B. Mills et al. (LSND Collaboration),
 Nucl. Phys. Proc. Suppl. {\bf 91}, 198 (2001).
 J. Kleinfeller et al. (KARMEN Collaboration),
Nucl. Phys. Proc. Suppl. {\bf 87}, 281 (2000). 
\bibitem{double}
H.V. Klapdor-Kleingrothaus, A. Dietz, H.L. Harney, and 
I.V. Krivosheina, Mod. Phys. Lett. {\bf A16}, 2409 (2001).

\bibitem{alter}
For a review, see G. Altarell, hep-ph/0206077. 
\bibitem{bar1}
M.~A.~Luty, Phys.\ Rev.\ D {\bf 45}, 455 (1992);
H.~Murayama and T.~Yanagida,Phys.\ Lett.\ B {\bf 322}, 349 (1994);
M.~Flanz, E.~A.~Paschos and U.~Sarkar, Phys.\ Lett.\ B {\bf 345}, 248 (1995)
[Erratum-ibid.\ B {\bf 382}, 447 (1995)];
M.~Plumacher,
Z.\ Phys.\ C {\bf 74}, 549 (1997);
L.~Covi, E.~Roulet and F.~Vissani, Phys.\ Lett.\ B {\bf 384}, 169 (1996);
E.~Ma and U.~Sarkar,
Phys.\ Rev.\ Lett.\  {\bf 80}, 5716 (1998);
A.~Pilaftsis,
Int.\ J.\ Mod.\ Phys.\ A {\bf 14}, 1811 (1999);
A.~Riotto and M.~Trodden, Ann.\ Rev.\ Nucl.\ Part.\ Sci.\  {\bf 49}, 35 (1999);
J.~R.~Ellis, S.~Lola and D.~V.~Nanopoulos,
Phys.\ Lett.\ B {\bf 452}, 87 (1999);
W.~Buchmuller and M.~Plumacher,
Int.\ J.\ Mod.\ Phys.\ A {\bf 15}, 5047 (2000);
D.~Falcone and F.~Tramontano,
Phys.\ Lett.\ B {\bf 506}, 1 (2001);
H.~B.~Nielsen and Y.~Takanishi, Phys.\ Lett.\ B {\bf 507}, 241 (2001);
A.~S.~Joshipura, E.~A.~Paschos and W.~Rodejohann,
JHEP {\bf 0108}, 029 (2001);
B.~Brahmachari, E.~Ma and U.~Sarkar,
Phys.\ Lett.\ B {\bf 520}, 152 (2001);
M.~Hirsch and S.~F.~King,
Phys.\ Rev.\ D {\bf 64}, 113005 (2001);
F.~Buccella, D.~Falcone and F.~Tramontano,
Phys.\ Lett.\ B {\bf 524}, 241 (2002);
W.~Buchmuller and D.~Wyler,
Phys.\ Lett.\ B {\bf 521}, 291 (2001);
M.~S.~Berger and K.~Siyeon,
Phys.\ Rev.\ D {\bf 65}, 053019 (2002);
G.~C.~Branco, R.~Gonzalez Felipe, F.~R.~Joaquim and M.~N.~Rebelo,
hep-ph/0202030;
M.~Fujii, K.~Hamaguchi and T.~Yanagida, hep-ph/0202210;
S.~Davidson and A.~Ibarra, hep-ph/0202239;
E.~A.~Paschos,
hep-ph/0204137;
W.~Buchmuller, hep-ph/0204288.


\bibitem{cpt}
H.~Murayama and T.~Yanagida,
Phys.\ Lett.\ B {\bf 520}, 263 (2001);
G.~Barenboim, L.~Borissov, J.~Lykken and A.~Y.~Smirnov, hep-ph/0108199;
A.~d .Gouvea, hep-ph/0204077.

\bibitem{4nu}
G.~L.~Fogli, E.~Lisi and A.~Marrone,
Phys.\ Rev.\ D {\bf 63}, 053008 (2001);
O.~L.~Peres and A.~Y.~Smirnov, 
Nucl.\ Phys.\ B {\bf 599}, 3 (2001);
C.~Giunti, Nucl.\ Phys.\ Proc.\ Suppl.\  {\bf 100}, 244 (2001);
W.~Grimus and T.~Schwetz, Eur.\ Phys.\ J.\ C {\bf 20}, 1 (2001);
M.~C.~Gonzalez-Garcia, M.~Maltoni and C.~Pena-Garay, hep-ph/0108073;
M.~Maltoni, T.~Schwetz and J.~W.~Valle, hep-ph/0112103;
A.~Strumia, hep-ph/0201134.
R. Barbieri et al., JHEP {\bf 9812}, 017 (1998) and references therein;
S.P. Mikheyev and A. Yu. Smirnov, Sov. J. Nucl. Phys. {\bf 42}, 913 (1986);
G.L. Fogli, E. Lisi, D. Montanino, Phys. Rev. {\bf D54}, 2048 (1996);
S. Bilenky, C. Giunti, and C. Kim, Astrop. Phys. {\bf 4}, 241 (1996);
T. Teshima, T. Sakai, and O.Ingaki, Int. Mod. Phys. {\bf A14}, 1953 (1999);
M.C. Gonzalez-Garcia, P.C. de Holanda, C. Pena-Garay, and J.W.F. Valle,
hep-ph/9906469;
J.N. Bahcall, P.I. Krastev, and A.Yu. Smirnov, Phys. Rev. {\bf D58}, 
096016 (1998);
E. Malkawi, Phys. Rev. {\bf D61}, 013006 (2000).
S.M. Bilenky and C. Giunti, hep-ph/9905246;
S.M. Bilenky, C. Giunti, W Grimus, and T. Schwetz, hep-ph/9904316.
Yutaka Okamoto and Masaki Yasue, Prog. Theor. Phys. 
{\bf 101}, 1119 (1999);
N. Gaur, A. Ghosal, E. Ma, and P. Roy, Phys. Rev. {\bf D58}, 071301 (1998).


\bibitem{nuexd}
K.~R.~Dienes, E.~Dudas and T.~Gherghetta,
Nucl.\ Phys.\ B {\bf 557}, 25 (1999);
N.~Arkani-Hamed, S.~Dimopoulos, G.~Dvali and J.~March-Russell,
Phys.\ Rev.\ D {\bf 65}, 024032 (2002).

\bibitem{stab} .
K.~S.~Babu, E.~Ma and J.~W.~Valle, hep-ph/0206292;
E.~Ma, Mod.\ Phys.\ Lett.\ A {\bf 17}, 627 (2002);
J.R.~Ellis, S.~Lola, Phys. Lett. {\bf  B458}, 310 (1999);
N.~Haba, Y.~Matsui, N.~Okamura, M.~Sugiura, Eur.\ Phys.\ J. C10, 677 (1999);
J.A.~Casas, J.R.~Espinosa, A.~Ibarra, I.~Navarro, Nucl. Phys. 
{\bf B569}, 82 (2000);
JHEP 9909, 015 (1999);
Nucl. Phys. {\bf B573}, 652 (2000);
N.~Haba, N.~Okamura, Eur.\ Phys.\ J. C14, 347 (2000);
P.H.~Chankowski, W.~Krolikowski, S.~Pokorski, Phys. Lett. {\bf B473}
, 109 (2000);
T.~K.~Kuo, J.~Pantaleone and G.~H.~Wu,
Phys.\ Lett.\ B {\bf 518}, 101 (2001);
P.~H.~Chankowski and S.~Pokorski,
Int.\ J.\ Mod.\ Phys.\ A {\bf 17}, 575 (2002);
S.~Antusch, J.~Kersten, M.~Lindner and M.~Ratz,


\bibitem{ehab2}  
H. Widyan, E. Malkawi, and M.B. Altai. Accepted for publication in Hadronic
Journal;
E. Malkawi, published in the proceedings of the  
Cairo Int. Conf. on High Energy Physics (January, 9-14, 2001),  
ed. S. Khalil, Q. Shafi, \& H. Tallat.

 \bibitem{ehab1}  E. Malkawi, T. Tait, and C.--P. Yuan. Phys. Lett. {\bf B385},
 304 (1996); 
  D.J. Muller and S. Nandi, Phys. Lett. {\bf B383}, 345 (1996);
  K.Y. Lee and J.C. Lee, Phys. Rev. {\bf D58}, 115001 (1998);
   J.C. Lee, K.Y. Lee, and J.K. Kim,  Phys. Lett. {\bf B424}, 133 (1998);
E. Malkawi and C.-P. Yuan, Mod. Phys. Lett. {\bf A14}, 1487 (1999);
E. Malkawi and C.-P. Yuan, Phys. Rev. {\bf D61}, 015007 (2000).

\bibitem{lima}
 X.Y. Li and E. Ma, Phys. Rev. Lett. {\bf 47},
 1788 (1981); E. Ma,  X. Li, and S.F. Tuan
 Phys. Rev. Lett. {\bf 60},
 495 (1988); X. Li and E. Ma, Phys. Rev. {\bf D46}, 1905 (1992);
 J. Phys. {\bf G19}, 1265 (1993).


\bibitem{data}
Review of Particle Physics, The European Physical Journal {\bf C15}, 1 (2000).

\bibitem{smir}
V. Barger et al., Phys. Lett. {\bf B 537}, 179 (2002);
J.N. Bahcall, P.I. Krastev, A.Yu. Smirnov, JHEP {\bf 0105}, 015 (2001);
P.C. de Holanda, A.Yu. Smirnov, hep-ph/0205241;
P.I. Krastev, A.Yu. Smirnov, Phys. Rev. {\bf D65}, 073022 (2002);
J.N. Bahcall, M.C. Gonzalez-Garcia, C. Pena-Garay, 
JHEP {\bf 0108}, 014 (2001); 
C.~V.K.~Baba, D.~Indumathi and M.~V.N.~Murthy, 
Phys.\ Rev.\ D {\bf 65}, 073033 (2002);
A.M. Gago et al., Phys.Rev. {\bf D65}, 073012 (2002);
V. Berezinsky, Astropart. Phys. {\bf 17}, (509) 2002.  

\bibitem{dark}
Xiang-dong Shi, George M. Fuller,  
Phys. Rev. Lett. {\bf 82}, 2832 (1999);
Alexander Kusenko, Gino Segre, Phys. Lett. {\bf B396}, 197 (1997). 
\end{thebibliography}
\end{document}